\documentclass[aps,prd,amssymb,cite,
amsfonts,epsf,preprintnumbers,nofootinbib,superscriptaddress]{revtex4}

\usepackage[dvips]{graphicx}
\usepackage{bm,latexsym,amsmath,amssymb,amsfonts}
\usepackage[usenames,dvipsnames]{color}
\usepackage[colorlinks=true,linkcolor=blue]{hyperref}
\usepackage{color}
\usepackage{soul}
\soulregister\cite7
\soulregister\ref7
\soulregister\pageref7
\usepackage{epsfig}
\usepackage{braket}
\usepackage[mathscr]{eucal}
\usepackage{cancel}
\usepackage{mathrsfs}
\usepackage{pgf, tikz}
\usepackage{slashed}
\usetikzlibrary{arrows,automata}
\usepackage{natbib}
\usepackage{pgfplots}
\pgfplotsset{compat=newest}

\definecolor{mypink1}{rgb}{0.858, 0.188, 0.478}
\definecolor{mypink2}{RGB}{219, 48, 122}
\definecolor{mypink3}{cmyk}{0, 0.7808, 0.4429, 0.1412}
\definecolor{mygray}{gray}{0.6}
\definecolor{pptbg}{rgb}{0.961,0.945,0.863}

\newcommand{\be}[1]{\begin{equation} \label{#1}}
\newcommand{\ee}{\end{equation}}
\newcommand{\bea}{\begin{eqnarray}}
\newcommand{\bean}{\begin{eqnarray*}}
\newcommand{\eea}{\end{eqnarray}}
\newcommand{\eean}{\end{eqnarray*}}
\newcommand{\ba}{\begin{array}}
\newcommand{\ea}{\end{array}}
\newcommand{\bel}{\begin{align}}
\newcommand{\eel}{\end{align}}
\newcommand{\nn}{\nonumber}

\newcommand{\qarrow}{\quad\rightarrow\quad}

\newcommand{\doi}[1]{\href{#1}{\color{blue}{#1}} }

\newcommand{\cS}{\mathcal{S}}

\begin{document}
\title{A master equation for Carter-separable stationary axisymmetric spacetimes and compatible sources}
\author{Hyeong-Chan Kim}
\affiliation{School of Liberal Arts and Sciences, Korea National University of Transportation, Chungju 380-702, Korea}

\email{hckim@ut.ac.kr}

\begin{abstract}
We show that the remaining diagonal Einstein equations in the
Carter-projective sector of stationary axisymmetric spacetimes are
equivalent to a single sourced master equation.
The projective structure is taken as the input fixed by the off-diagonal Einstein equations.  
In the anti-aligned exponential branch, which contains the Kerr--Carter and Pleba\'nski--Demia\'nski real section, the remaining diagonal Einstein system reduces to
\[
 \mathcal L_{\rm CP}[\Delta,Y]
    =16\pi\Sigma \left(  T_{\hat0\hat0}+T_{\hat3\hat3} \right),
\]
where \(\Delta(r)\) and \(Y(x)\) are the radial and angular structure
functions.  
The reduction is accompanied by two geometric diagonal identities of the Einstein tensor, which become algebraic compatibility conditions on admissible matter sources.  
In the homogeneous limit, the vacuum--\(\Lambda\) Kerr--Carter and Pleba\'nski--Demia\'nski families are recovered as solutions of the same master operator.  
We also show the projective covariance of the construction and discuss compatible sources, including the aligned Maxwell field and separable anisotropic examples.
\end{abstract}
\maketitle

\section{Introduction} \label{sec:intro}

Stationary and axisymmetric spacetimes provide one of the most important arenas in which exact solutions of the Einstein equations can be constructed, classified, and related to black-hole uniqueness results~\cite{Carter71,Robinson1975,Mazur2000,Chrusciel2012}.
In its most general form, a stationary axisymmetric metric may be written in the Weyl--Lewis--Papapetrou form~\cite{Weyl1917,Lewis1932,Papapetrou1953,Chandrasekhar83},
\be{Papapetrou}
ds^2
= -f(\rho,z)\left(dt-\omega(\rho,z)d\phi\right)^2 +f^{-1}(\rho,z)
\left[ e^{2\gamma(\rho,z)}(d\rho^2+dz^2)+\rho^2d\phi^2
\right].
\ee
The corresponding Einstein equations are naturally formulated as the Ernst system~\cite{Ernst1968,Ernst1968b}.
This framework is well suited to the general stationary axisymmetric problem, but it is broader than the class of rotating geometries possessing Carter-type separability, such as the Kerr, Carter, and Pleba\'nski--Demia\'nski families~\cite{Kerr1963,Carter1968,Plebanski76}.

The central result of this paper is that, once the Carter-projective
structure is fixed by the off-diagonal Einstein equations, the remaining diagonal system is equivalent to a {\it single sourced master equation} for the radial and angular structure functions, $\Delta(r)$ and $Y(x)$, together with algebraic compatibility conditions on matter.
 In the homogeneous limit this master equation reproduces the
Carter and Pleba\'nski--Demia\'nski quartic structure functions, while in the sourced case it organizes the admissible matter deformations through the single combination $T_{\hat0\hat0}+T_{\hat3\hat3}$.

These separable geometries possess a rigid radial--angular
organization~\cite{Carter1968,Chandrasekhar83}, closely tied to hidden symmetries and to the algebraically special structure of rotating black-hole
solutions~\cite{Newman1962,Goldberg62,WalkerPenrose1970}.  The Kerr--Carter and Pleba\'nski--Demia\'nski families are four-dimensional type-D separable geometries admitting hidden symmetries, such as Killing--Yano or conformal Killing--Yano structures~\cite{WalkerPenrose1970,DietzRuediger1981}.  Related modern developments in hidden symmetries and separability are reviewed in the Kerr--NUT--AdS context in Refs.~\cite{FrolovKubiznak2007,Kubiznak2007,Frolov2007}.

We use the Einstein equation in the convention
\be{Einstein-eq}
 G_{\hat a\hat b}+\Lambda\eta_{\hat a\hat b}
 =
 8\pi T_{\hat a\hat b}.
\ee
The Carter-type separable sector considered below relates  the radial and angular variables by a projective structure and combines the two Killing directions into the separated one-forms characteristic of Carter separability. 
This restriction excludes generic Weyl multipole data, but it makes the field equations highly constrained and reorganizes them into equations for a small number of radial and angular structure functions.

In previous work~\cite{Kim:2026gog,Kim:2026wkk}, the off-diagonal Einstein equations were shown to fix the underlying Carter-projective structure.  
In particular, after introducing characteristic variables $R$ and $z$, the equation $G_{\hat0\hat3}=0$ determines the conformal factor up to a function of $y=R-z$, while $G_{\hat1\hat2}=0$ gives a Riccati equation for this remaining function.  
In the generic closure branch, this Riccati equation reduces to an ordinary differential equation in $y$, leading to a common Schwarzian condition for the radial and angular projective functions and to a projective alignment between the two sectors,
reflecting the underlying projective geometry~\cite{Lehto87}.  
In the present paper we take this projective reduction as the input and ask how the remaining diagonal Einstein equations are organized.

We focus on  the anti-aligned exponential branch, which contains the Lorentzian Kerr--Carter and Pleba\'nski--Demia\'nski real section~\cite{Carter1968,Plebanski76}.
After the off-diagonal equations have fixed the Carter-projective geometry, the Einstein tensor satisfies the geometric identities
\be{intro-identities}
G_{\hat0\hat0}+G_{\hat1\hat1}=0,
\qquad
G_{\hat2\hat2}-G_{\hat3\hat3}=0.
\ee
Thus these two diagonal combinations do not generate further structure equations.  
Instead, in the presence of matter, they become algebraic compatibility conditions on the admissible stress tensor. 
In operator form, the remaining independent diagonal combination gives
\be{master1}
  \mathcal L_{\rm CP}[\Delta,Y]
 = 16\pi\Sigma
 \left(
 T_{\hat0\hat0}+T_{\hat3\hat3}
 \right),
\ee
where $\Delta=\Delta(r)$ and $Y(x)=(1-x^2)Q(x)$.  
The operator $\mathcal L_{\rm CP}$ is completely determined by the projective data and by the Carter-compatible conformal factor.  
We refer to \eqref{master1} as the master equation of the Carter-projective sector.

This master-equation viewpoint gives a unified interpretation of the
Kerr--Carter and Pleba\'nski--Demia\'nski families.  
When the compatible source vanishes, the equation becomes homogeneous.  
Its non-accelerating representative yields the Carter quartic structure functions~\cite{Carter1968}, while the accelerating representative yields the Pleba\'nski--Demia\'nski quartic structure~\cite{Plebanski76}.  
Thus these familiar vacuum--$\Lambda$ solutions arise as distinguished homogeneous solutions of the same Carter-projective master operator, with the acceleration dependence entering through the allowed conformal factor.

The sourced equation also clarifies which matter configurations can preserve the same separable structure.  
The identities~\eqref{intro-identities} require $ T_{\hat0\hat0}+T_{\hat1\hat1}=0$ and $ T_{\hat2\hat2}-T_{\hat3\hat3}=0$. 
For a stress tensor diagonal in the Carter-projective orthonormal
frame, these conditions become
\be{compatibility-intro}
 p_r=-\rho,
 \qquad
 p_\theta=p_\phi .
\ee
This vacuum-like radial equation of state frequently appears in regular rotating black-hole models based on de Sitter-type cores~\cite{Dymnikova1992,Hayward2006}, nonlinear electrodynamics sources~\cite{AyonBeatoGarcia1998,AyonBeatoGarcia1999,Toshmatov2017}, and Kerr--Schild or Newman--Janis constructions of rotating regular geometries~\cite{GursesGursey1975,BambiModesto2013, DymnikovaGalaktionov2015}.  
In the present formulation, however, these conditions are \emph{not imposed by hand}; they follow from compatibility with the Carter-projective sector, while the remaining freedom enters through the single source term in the master equation.
Thus the sourced master equation is the organizing object for both the vacuum Carter--Pleba\'nski--Demia\'nski sector and its compatible matter deformations.

We also show that the construction is projectively covariant. 
A common M\"obius transformation of the radial and angular projective variables preserves the projective invariant controlling the conformal structure.
With the corresponding projective weights of $\Delta$ and $Y$, both the metric ansatz and the master equation retain their form. 
Therefore the simple representative used in the explicit derivation is a projective gauge choice rather than a loss of generality.

The paper is organized as follows. In Sec.~\ref{sec:separable} we introduce
the Carter-type separable ansatz and summarize the previous off-diagonal
projective reduction. In Sec.~\ref{sec:anti-exp} we specialize to the
anti-aligned exponential branch relevant for the Kerr--Carter and
Pleba\'nski--Demia\'nski real section. In Sec.~\ref{sec:master} we derive
the master equation from the remaining diagonal Einstein equation. In
Sec.~\ref{sec:exact} we recover the Carter and Pleba\'nski--Demia\'nski
quartic structure functions in the homogeneous limit. In
Sec.~\ref{sec:covariance} we discuss projective covariance, and in
Sec.~\ref{sec:matter} we describe compatible matter sources.

\section{Carter-projective ansatz and previous reduction}
\label{sec:separable}

In this section we introduce the Carter-type separable ansatz used in this work and summarize the projective reduction obtained from the off-diagonal Einstein equations. 
The only difference is that the starting metric~\eqref{metric:gen} is a bit more general than that in the previous work, which is a gauge artifact. 
The purpose of this section is not to repeat the full derivation of the projective classification, but to fix notation and to state the results that will be used in the analysis of the remaining diagonal equation.

\subsection{Carter-type separable ansatz and characteristic variables}
\label{subsec:ansatz}

Guided by the radial/angular separability of the Kerr--Carter and
Pleba\'nski--Demia\'nski geometries, we consider the stationary axisymmetric
metric ansatz
\be{metric:gen}
ds^2 =
 -  \frac{\Sigma \Delta}
	{q\left(\Gamma - a^2p \right)^2}
	\left( dt - a p d\phi \right)^2
+ \frac{\Sigma}{q\Delta }dr^2
+ \frac{\Sigma}{Q\Xi} d\theta^2
+ \frac{\Sigma Q \sin^2\theta}
	{\Xi \left(\Gamma - a^2p \right)^2}
	(\Gamma d\phi -a dt)^2 ,
\ee
where
\bea
\Delta=\Delta(r),\qquad
\Gamma=\Gamma(r),\qquad
\Xi=\Xi(r),
\nn\\
p=p(x),\qquad
q=q(x),\qquad
Q=Q(x),\qquad
\Sigma=\Sigma(r,x),
\eea
with
\be{x-def}
x\equiv \cos\theta .
\ee
The two Killing directions are arranged into the separated one-forms
\[
 dt-ap(x)d\phi,
 \qquad
 \Gamma(r)d\phi-a\,dt .
\]
Thus the ansatz is not a rewriting of the most general
Weyl--Lewis--Papapetrou metric. It is a restriction to a Carter-type
separable sector in which the radial and angular dependences are organized
by a projective structure.

The functions \(q(x)\) and \(\Xi(r)\) can be removed locally. First, one
absorbs the product \(q\Xi\) into the conformal factor by redefining
\[
 \frac{\Sigma}{q\Xi}\longrightarrow \Sigma .
\]
Then the radial and angular coordinates may be reparametrized locally so
that
\[
 \Xi\,dr\longrightarrow dr,
 \qquad
 q\,dx\longrightarrow dx .
\]
Therefore, in what follows we work in the representative
\be{qXi-gauge}
q=1,
\qquad
\Xi=1.
\ee
The metric becomes
\be{metric:gen2}
ds^2 =
 -  \frac{\Sigma \Delta}
	{\left(\Gamma - a^2p \right)^2}
	\left( dt - a p d\phi \right)^2
+ \frac{\Sigma}{\Delta }dr^2
+ \frac{\Sigma }{Q\sin^2\theta } dx^2
+ \frac{\Sigma Q \sin^2\theta}
	{\left(\Gamma - a^2p \right)^2}
	(\Gamma d\phi -a dt)^2 .
\ee
For later use, we define the angular structure function
\be{Y-def}
Y(x)\equiv (1-x^2)Q(x).
\ee

The Pleba\'nski--Demia\'nski acceleration factor is not excluded by the
ansatz. Rather, if an accelerating representative is present, the
acceleration dependence must appear through the Carter-compatible conformal
factor \(\Sigma\). In this sense the ansatz captures the accelerating
sector only when the conformal factor is compatible with the same
radial/angular projective alignment.

\vspace{.3cm}
We introduce characteristic variables \(R\) and \(z\) by
\be{Rz:rx}
\partial_R \equiv \Gamma'(r)\,\partial_r,
\qquad
\partial_z \equiv \dot p(x)\,\partial_x .
\ee
Here a prime denotes \(d/dr\), while a dot denotes \(d/dx\). Equivalently,
\[
 \Gamma_R=(\Gamma')^2,
 \qquad
 p_z=(\dot p)^2 .
\]
The detailed relation between \((R,z)\) and \((r,x)\) depends on the
projective branch and will be discussed in App.~\ref{app:coordinates}. We
also define
\be{y}
 y\equiv R-z,
 \qquad
 \bar y\equiv R+z .
\ee
The variable \(y\) plays the role of the characteristic coordinate along
which the conformal factor retains one functional degree of freedom.

\subsection{Summary of the previous projective reduction}
\label{subsec:previous-reduction}

We now summarize the part of the Einstein equations that fixes the
Carter-projective sector, the main results in Ref.~\cite{Kim:2026wkk}. 
The off-diagonal equation \(G_{\hat0\hat3}=0\), written in the orthonormal frame adapted to \eqref{metric:gen2}, determines
the conformal factor as
\be{Sigma}
\Sigma(R,z)
=
\frac{(\Gamma-a^2p)\varSigma(y)}
{\sqrt{\Gamma_Rp_z}} .
\ee
Equivalently, defining
\be{S:Sigma}
S(R,z)
\equiv
\frac{(\Gamma-a^2p)^2}{\Gamma_Rp_z},
\ee
one may write
\be{Sigma-S}
\Sigma(R,z)
=
\epsilon\sqrt{S(R,z)}\,\varSigma(y),
\qquad
\epsilon=\pm1 .
\ee

The second off-diagonal equation \(G_{\hat1\hat2}=0\) becomes a Riccati
equation for
\be{sigma-def}
\sigma(y)\equiv \frac{\mathring{\varSigma}}{\varSigma},
\qquad
\mathring{\sigma}\equiv \frac{d\sigma}{dy}.
\ee
Its form is
\be{Riccati}
\mathring \sigma
=
\frac12\sigma^2
+\frac12F(R,z)\sigma
+\frac14G(R,z),
\ee
where
\bea
F(R,z)
&=&
\frac{\Gamma_R+a^2p_z}{\Gamma-a^2p}
+\frac12\,\partial_z\log p_z
-\frac12\,\partial_R\log\Gamma_R ,
\nn\\
G(R,z)
&=&
-\frac{\Gamma_{RR}}{\Gamma_R}
 \frac{a^2p_z}{\Gamma-a^2p}
+\frac{p_{zz}}{p_z}
 \frac{\Gamma_R}{\Gamma-a^2p}
-\frac12
 \frac{\Gamma_{RR}}{\Gamma_R}
 \frac{p_{zz}}{p_z}.
\label{FG-def}
\eea
The full derivation is reviewed in App.~\ref{app:pre}.

In the generic closure branch, the Riccati equation is required to be an
ordinary differential equation in the single variable \(y\). Thus
\be{closure}
F(R,z)=F(y),
\qquad
G(R,z)=G(y).
\ee
This condition leads to the equality of the Schwarzian derivatives,
\be{SchD}
\{\Gamma,R\}=\{p,z\}=C,
\qquad
\{A,B\}\equiv
\frac{A_{BBB}}{A_B}
-\frac32
\left(\frac{A_{BB}}{A_B}\right)^2 .
\ee
Hence \(\Gamma(R)\) and \(p(z)\) are fixed up to projective
transformations. The local representatives are classified by the sign of
the common Schwarzian constant:
\be{GammaR}
\Gamma(R)=
\begin{cases}
\dfrac{\alpha+\beta R}{\gamma+\delta R},
& C=0,\\[8pt]
\dfrac{\alpha e^{\mu R}+\beta e^{-\mu R}}
{\gamma e^{\mu R}+\delta e^{-\mu R}},
& C=-2\mu^2,\\[8pt]
\dfrac{\alpha\cos(\mu R)+\beta\sin(\mu R)}
{\gamma\cos(\mu R)+\delta\sin(\mu R)},
& C=+2\mu^2,
\end{cases}
\ee
with
$
 \alpha\delta-\beta\gamma\neq0,
$
and with an analogous expression for \(p(z)\).

The closure condition imposes a further alignment between the radial and
angular projective variables. In the generic branch one obtains
\be{pGamma}
p(z)
=
\frac1{a^2}\,
\Gamma\!\left(\pm(z-z_0)\right),
\ee
where \(z_0\) is a constant. We refer to this relation as the
\emph{projective alignment} of the radial and angular sectors. The two
signs correspond to the aligned and anti-aligned choices along the
characteristic directions.

In the present work we focus on the anti-aligned exponential branch,
because this branch contains the Lorentzian Kerr--Carter and
Pleba\'nski--Demia\'nski real section. The projective classification above
corresponds to the generic closure branch. A non-generic cancellation
branch is logically possible, in which \(F\) and \(G\) are not separately
functions of \(y\), but the particular combination appearing in
\eqref{Riccati} is. Such a branch requires additional overdetermined
compatibility conditions. Since the purpose of this paper is to analyze the
remaining diagonal Einstein equation in the Pleba\'nski--Demia\'nski
containing branch, we do not classify such isolated possibilities here.

\section{Anti-aligned exponential branch}
\label{sec:anti-exp}

In the rest of this work we focus on the anti-aligned exponential branch.
This is the projective branch that contains the Lorentzian Kerr--Carter and
Pleba\'nski--Demia\'nski real section.  In this section we collect the
explicit branch data and introduce the representative that will be used in
the derivation of the master equation.

\subsection{Branch data and conformal factor}
\label{subsec:anti-exp-data}

The anti-aligned exponential branch is described by
\be{GammaR:anti}
\Gamma(R)=
\frac{\alpha e^{\mu R}+\beta e^{-\mu R}}
{\gamma e^{\mu R}+\delta e^{-\mu R}},
\qquad
p(z)=\frac1{a^2}
\frac{\alpha e^{-\mu z_\ominus}-\beta e^{\mu z_\ominus}}
{\gamma e^{-\mu z_\ominus}-\delta e^{\mu z_\ominus}},
\qquad
z_\ominus\equiv z-z_r ,
\ee
where \(z_r\) is a real constant.  This form is obtained from the
anti-aligned relation by taking the shifted representative
\be{z0}
 z_0=z_r+\frac{\pi i}{2\mu}.
\ee
In this patch one has
\be{GammaR-pz-anti}
\Gamma_R=
\frac{2\mu D_0}
{\left(\gamma e^{\mu R}+\delta e^{-\mu R}\right)^2},
\qquad
p_z=
\frac{2\mu D_0}
{a^2\left(\gamma e^{-\mu z_\ominus}
-\delta e^{\mu z_\ominus}\right)^2},
\ee
where
$ 
D_0\equiv \alpha\delta-\beta\gamma .
$
Thus \(\Gamma_R\geq0\) and \(p_z\geq0\) are ensured by choosing
\[
 \frac{\mu D_0}{a^2}>0 .
\]

For this branch the projective invariant \(S\) defined in
\eqref{S:Sigma} becomes
\be{S-anti-exp}
S(R,z)=
\frac{a^2}{\mu^2}
\cosh^2[\mu(\bar y-z_r)] .
\ee
The Riccati equation is solved by
\be{varSigma}
\varSigma(y)=
\frac{\varSigma_0}
{\left(
Ae^{\mu(y+z_r)/2}
+
Be^{-\mu(y+z_r)/2}
\right)^2}.
\ee
Therefore, using \(\Sigma=\epsilon\sqrt{S}\,\varSigma(y)\), we obtain
\be{Sigma:anti-exp}
\Sigma(R,z)
=
\frac{\epsilon a\varSigma_0}{\mu}
\frac{\cosh[\mu(\bar y-z_r)]}
{\left(
Ae^{\mu(y+z_r)/2}
+
Be^{-\mu(y+z_r)/2}
\right)^2},
\qquad
\epsilon=\pm1 .
\ee
The branch data above solve the two off-diagonal equations
\(G_{\hat0\hat3}=0\) and \(G_{\hat1\hat2}=0\) as shown in App.~\ref{app:checks}.  

\subsection{Kerr--Carter/Pleba\'nski--Demia\'nski representative}
\label{subsec:Kerr-PD-representative}

The full relation between the projective coordinates \((R,z)\) and the
coordinates \((r,x)\) is branch-dependent and is summarized in
App.~\ref{app:coordinates2}.  For the Kerr--Carter and
Pleba\'nski--Demia\'nski real section, it is sufficient to choose the
projective representative
\[
 \gamma=0 .
\]
After shifts and rescalings of \(r\) and \(x\), one may write
\be{R:r}
e^{\mu R}
=
\epsilon
\sqrt{\frac{\mu\delta}{2\alpha}}\,r,
\qquad
e^{\mu z_\ominus}
=
\epsilon
\sqrt{\frac{2\alpha}{\delta\mu}}\,
\frac1{a x_-},
\ee
where
\[
 x_-\equiv x-x_0 .
\]
In the same representative,
\be{Gamma-p-representative}
\Gamma(r)
=
\Gamma_0+\frac{\mu}{2}r^2,
\qquad
a^2p(x)
=
\Gamma_0-\frac{\mu}{2}a^2x_-^2,
\qquad
\Gamma_0\equiv \frac{\beta}{\delta}.
\ee
Therefore,
\be{Gamma-p}
\Gamma-a^2p
=
\frac{\mu}{2}
\left(r^2+a^2x_-^2\right).
\ee
This is the characteristic Kerr--Carter sum structure.  The
non-accelerating Carter representative and the accelerating
Pleba\'nski--Demia\'nski representative will arise from different choices
of the remaining conformal factor parameter in \(\varSigma(y)\), namely
from the cases \(A=0\) and \(A\neq0\), respectively.

\subsection{Automatic diagonal identities and matter compatibility}

A useful property of the anti-aligned exponential branch is that, after the
off-diagonal equations have fixed the Carter-projective sector, the
Einstein tensor satisfies the two diagonal identities, as shown in App.~\ref{app:diagonal-eqs},
\begin{equation}
G_{\hat0\hat0}+G_{\hat1\hat1}=0,
\qquad
G_{\hat2\hat2}-G_{\hat3\hat3}=0.
\label{automatic-diagonal-identities}
\end{equation}
These relations should not be interpreted as consequences of the
vacuum--\(\Lambda\) Einstein equations.  
Rather, they are geometric identities of the Kerr/Carter-projective sector itself.  
In other words, once the off-diagonal equations have selected this sector, the corresponding Einstein tensor has fewer independent diagonal components than a generic stationary axisymmetric geometry.

This observation becomes important when matter is included.
We take the Einstein equation in the form~\eqref{Einstein-eq}.
Notice that the cosmological constant drops out of the two combinations in
\eqref{automatic-diagonal-identities}.  
Therefore, any matter source compatible with the same Kerr/Carter-projective geometry must satisfy
\begin{equation}
T_{\hat0\hat0}+T_{\hat1\hat1}=0,
\qquad
T_{\hat2\hat2}-T_{\hat3\hat3}=0.
\label{matter-compatibility-conditions}
\end{equation}
These are not additional gravitational field equations for the metric
functions.  They are algebraic compatibility conditions on the admissible
stress tensor.
The remaining diagonal combination $G_{\hat0\hat0}+G_{\hat3\hat3}$ then carries the only independent matter information; this is the master equation
derived in Sec.~\ref{sec:master}, and the resulting class of compatible
sources is analyzed in Sec.~\ref{sec:matter}.

\section{The master equation} \label{sec:master}

The projective classification is determined by the off-diagonal Einstein
equations, $G_{\hat0\hat3}=0,$ and $ G_{\hat1\hat2}=0.$
Once these equations are imposed, the anti-aligned exponential
Carter-projective sector satisfies the diagonal identities~\eqref{automatic-diagonal-identities}.
After these conditions are imposed, the remaining independent matter
information enters through~\eqref{master1}.
This equation gives the master equation for the radial and angular
structure functions.

\subsection{General form}

We now evaluate Eq.~\eqref{master1} in the Carter-projective orthonormal
frame.  A direct calculation gives, as shown in App.~\ref{app:master},
\be{G00+G33-1}
2\Sigma
\left(
G_{\hat0\hat0}+G_{\hat3\hat3}
\right)
=
\Delta''
+
(\partial_r\mathcal{C}_{\Delta'})\Delta'
+
\mathcal{C}_\Delta(\Delta-a^2Y)
-
(\partial_x\mathcal{C}_{\Delta'})\dot Y
-
\ddot Y
=
\cS(r,x),
\ee
where
\bea \label{CDelta1}
\mathcal{C}_\Delta(r,x)
&=&
\frac{2}{\Gamma-a^2p}
\partial_y
\left(
\log
\frac{(\Gamma-a^2p)^2}{\varSigma(y)}
\right),
\nn \\
\mathcal{C}_{\Delta'}(r,x)
&=&
\log
\frac{\Sigma}{(\Gamma-a^2p)^3}
=
\log
\frac{\varSigma}
{\sqrt{\Gamma_Rp_z}(\Gamma-a^2p)^2}.
\eea
The source term is
\be{source-term}
\cS(r,x)
=
16\pi\Sigma
\left(
T_{\hat0\hat0}+T_{\hat3\hat3}
\right).
\ee
Equation~\eqref{G00+G33-1} is the master equation of the
Carter-projective sector.  In the homogeneous case,
\begin{equation}
 \cS(r,x)=0,
\end{equation}
it reduces to the equation whose solutions give the Kerr--Carter and
Pleba\'nski--Demia\'nski structure functions.  For compatible matter
sources, the same geometric operator is sourced only by the single
combination \(T_{\hat0\hat0}+T_{\hat3\hat3}\).

It is useful to write the same equation in operator form.  Defining
\begin{equation}
 \mathcal M(r,x)
 \equiv
 e^{\mathcal C_{\Delta'}},
\end{equation}
Eq.~\eqref{G00+G33-1} becomes
\be{master-divergence-form}
\partial_r(\mathcal M\Delta')
-
\partial_x(\mathcal M\dot Y)
+
\mathcal M\mathcal C_\Delta(\Delta-a^2Y)
=
\mathcal M\cS .
\ee
Equivalently, introducing
\begin{equation}
 \Psi(r,x)
 \equiv
 \Delta(r)-a^2Y(x),
\end{equation}
one may write
\be{master-Psi-form}
\frac{1}{\mathcal M}\left[\partial_r(\mathcal M\partial_r\Psi)
+\frac1{a^2}\partial_x(\mathcal M\partial_x\Psi) \right]
+\mathcal C_\Delta\Psi =\cS .
\ee
Thus the master equation can be viewed as the restriction of the two-variable
linear equation \eqref{master-Psi-form} to the additively separated form
\(\Psi(r,x)=\Delta(r)-a^2Y(x)\).

\subsection{Independence of the remaining diagonal equations}
\label{subsec:independent-diagonal}

Let us comment on why the single combination
\(G_{\hat0\hat0}+G_{\hat3\hat3}\) is sufficient for determining the
remaining structure functions in the reduced Carter-projective system.
After the off-diagonal equations have fixed the projective sector, the
anti-aligned exponential branch satisfies the two geometric identities
\be{diagonal-identities-master-section}
G_{\hat0\hat0}+G_{\hat1\hat1}=0,
\qquad
G_{\hat2\hat2}-G_{\hat3\hat3}=0 .
\ee
Therefore the diagonal part of the Einstein equations has fewer independent
components than in a generic stationary axisymmetric geometry.

In the vacuum--\(\Lambda\) case, the Einstein equations require
\[
G_{\hat0\hat0}=\Lambda,
\qquad
G_{\hat1\hat1}=-\Lambda,
\qquad
G_{\hat2\hat2}=-\Lambda,
\qquad
G_{\hat3\hat3}=-\Lambda .
\]
The two identities above already imply that the first pair and the second
pair are not independent.  The nontrivial equation that remains for the
structure functions is the sum
\[
G_{\hat0\hat0}+G_{\hat3\hat3}=0 ,
\]
or, in the presence of matter,
\[
G_{\hat0\hat0}+G_{\hat3\hat3}
=
8\pi
\left(
T_{\hat0\hat0}+T_{\hat3\hat3}
\right).
\]
This is precisely the master equation derived above.

The apparent remaining equations, for example \(G_{\hat0\hat0}=\Lambda\)
and \(G_{\hat3\hat3}=-\Lambda\) separately in the vacuum--\(\Lambda\) case,
do not give additional independent differential equations for
\(\Delta(r)\) and \(Y(x)\).  Once the off-diagonal equations, the automatic
diagonal identities, and the master equation are imposed, the residual
diagonal equation is fixed by the contracted Bianchi identity, up to the
constant parameters already appearing in the homogeneous solution.  In the
vacuum case this residual equation fixes the interpretation of the
integration constants, in particular the cosmological constant and the
normalization of the quartic coefficients, rather than introducing a new
structure equation.

\section{Vacuum limit and the Carter and Pleba\'nski--Demia\'nski structure functions}
\label{sec:exact}

In this section we solve the master equation in the presence of a cosmological constant but without additional source in the representative introduced in Sec.~\ref{sec:anti-exp}.  
We first discuss the non-accelerating limit, where the master equation reduces to the standard Carter form.  
We then keep the acceleration parameter in the conformal factor and recover the Pleba\'nski--Demia\'nski quartic structure.

\subsection{Non-accelerating Carter representative}
\label{sec:Kerr}

We first consider the non-accelerating representative.  In the
anti-aligned exponential branch, this corresponds to the choice
\be{Kerr-choice}
\gamma=0,
\qquad
A=0 .
\ee
Then the function \(\varSigma(y)\) in Eq.~\eqref{varSigma} becomes
\be{varsigma}
\varSigma(y)
=
\frac{\varSigma_0}{B^2e^{-\mu(y+z_r)}}
=
\frac{\varSigma_0}{B^2}
e^{\mu R}e^{-\mu(z-z_r)}
=
\bar\varSigma_0\,a x_- r ,
\qquad
\bar\varSigma_0
\equiv
\frac{\mu\delta}{2\alpha}
\frac{\varSigma_0}{B^2}.
\ee
Here we have used the representative relations in
Eq.~\eqref{R:r}.  The coefficient functions appearing in the master
equation are then
\bea
\mathcal C_\Delta(r,x)
&=&
\frac{4}{r^2+a^2x_-^2},
\nn\\
\mathcal C_{\Delta'}(r,x)
&=&
\log\frac{\varSigma}{r x_-}
-2\log(r^2+a^2x_-^2)
+\mbox{const.}
\label{C-functions-Kerr}
\eea

Introducing
\be{uG-def}
u\equiv ax_-,
\qquad
G(u)\equiv a^2Y(x),
\ee
and multiplying the master equation by \(r^2+u^2\), one obtains
\be{Kerreq}
(r^2+u^2)(\Delta''-G'')
-4r\Delta'
+4uG'
+4(\Delta-G)
=0 .
\ee
Here and below a prime denotes differentiation with respect to the argument
of the corresponding function, namely \(d/dr\) for \(\Delta(r)\) and
\(d/du\) for \(G(u)\).

The quartic structure follows directly from this equation.  Differentiating
\eqref{Kerreq} first with respect to \(r\) and then with respect to \(u\),
one finds
\be{Carter-third-derivative}
\frac{\Delta'''(r)}{r}
=
\frac{G'''(u)}{u}
=
\mbox{constant}.
\ee
Therefore \(\Delta(r)\) and \(G(u)\) are fourth-order polynomials with no
cubic terms.  Substitution back into \eqref{Kerreq} fixes the relative
quadratic and constant terms, giving
\be{Delta:Carter}
\Delta(r)
=
\frac{r^4}{\ell^2}
+\Delta_2 r^2
+\Delta_1 r
+\Delta_0,
\qquad
Y(x)
=
\frac{a^2x_-^4}{\ell^2}
-\Delta_2 x_-^2
+\frac{g_1}{a}x_-
+\frac{\Delta_0}{a^2}.
\ee
Here, $\Delta_1$ and $g_1$ correspond to the mass and nut charge, respectively.
This reproduces the Carter quartic structure functions.  
This non-accelerating representative is contained in the more general
accelerating branch discussed below; we have presented it separately only
because it provides the simplest transparent limit of the master equation.

\subsection{Accelerating representative and the Pleba\'nski--Demia\'nski family}
\label{sec:PD}

We now keep the acceleration parameter in the conformal factor.  In the
same projective representative \(\gamma=0\), this corresponds to
\be{PD-choice}
A\neq0 .
\ee
The coordinate relations, the functions \(\Gamma(r)\), \(p(x)\), and the
combination \(\Gamma-a^2p\) are the same as in the previous subsection.
However, the function \(\varSigma(y)\) now takes the form
\be{varsigma-PD}
\varSigma(y)
=
\bar\varSigma_0
\frac{a x_- r}
{\left(1+\bar A a x_- r\right)^2},
\qquad
\bar A
\equiv
\frac{A}{B}\frac{\mu\delta}{2\alpha}.
\ee
The coefficient functions in the master equation become
\bea
\mathcal C_\Delta(r,x)
&=&
\frac{4}{r^2+a^2x_-^2}
\left[
1+\frac{2\bar A a x_- r}{1+\bar A a x_-r}
\right],
\nn\\
\mathcal C_{\Delta'}(r,x)
&=&
\log\frac{\varSigma}{r x_-}
-2\log(r^2+a^2x_-^2)
+\mbox{const.}
\label{C-functions-PD}
\eea

Using again
\[
u=ax_-,
\qquad
G(u)=a^2Y(x),
\]
the master equation can be written as
\bea \label{de:PD}
D &\equiv&
(1+\bar A u r)
\left[
(r^2+u^2)(\Delta''-G'') -4r\Delta' +4uG' +4(\Delta-G)
\right] 
\nn \\
&&+2\bar A
\left[ (r^2+u^2)(rG'-u\Delta') +4ur(\Delta-G)
\right]
=0 .
\eea
Taking successive derivatives gives
\be{d4D}
\partial_r\partial_u\partial_r\partial_u D
=
2(1+3\bar A r u)
\left[
\Delta^{(4)}(r)-G^{(4)}(u)
\right]
=0 .
\ee
Therefore, away from the exceptional locus $1+3\bar A r u=0,$ the two structure functions satisfy
\be{quartic-condition-PD}
\Delta^{(4)}(r)=G^{(4)}(u).
\ee
Hence \(\Delta(r)\) and \(G(u)\) are quartic polynomials with the same quartic coefficient.

Substituting the general quartic ansatz into \eqref{de:PD} fixes the
relative coefficients.  One obtains
\bea
\Delta(r)
&=& \frac{r^4}{\ell^2} +\Delta_3 r^3 +\Delta_2 r^2 +\Delta_1 r +\Delta_0,
\nn\\
Y(x)
&=& \frac{a^2x_-^4}{\ell^2} +\bar A\Delta_1 a x_-^3
-\Delta_2 x_-^2 +\frac{\Delta_3}{\bar A a}x_- +\frac{\Delta_0}{a^2}.
\label{PDsol}
\eea
In this form, \(\Delta_1\) is the mass-type coefficient, while \(\Delta_3\) is the
acceleration-dressed NUT-type coefficient.
Especially, in the $\bar A \to 0$ limit, $\Delta_3 \to 0$ and the nut charge is reduced to the linear coefficient of $Y$.   
The NUT charge itself is most directly read off from the coefficient of the linear term in the angular polynomial \(a^2Y(x)\).
Together with the conformal factor \eqref{varsigma-PD}, this reproduces the
Pleba\'nski--Demia\'nski structure in the Carter-projective representative~\cite{Plebanski76,GriffithsPodolsky2009}.

\section{Projective covariance}\label{sec:covariance}

In this section we clarify the projective redundancy of the
Carter-projective representation.  This redundancy is important because the
representative used in the previous sections, in particular the choice
\(\gamma=0\), should not be interpreted as a physical restriction.  It is a
choice of projective gauge.

Consider a common M\"obius transformation of the radial and angular
projective variables,
\be{Mobius}
 \Gamma\longrightarrow \widetilde\Gamma
 =
 \frac{A\Gamma+B}{C\Gamma+D},
 \qquad
 a^2p\longrightarrow a^2\widetilde p
 =
 \frac{A(a^2p)+B}{C(a^2p)+D},
\ee
where
$ J\equiv AD-BC\neq0 .$
For later convenience, we define
\be{LrLx}
 L_r\equiv C\Gamma+D,
 \qquad
 L_x\equiv Ca^2p+D .
\ee
Then
\be{tilde-Gamma-p}
 \widetilde\Gamma-a^2\widetilde p
 =
 \frac{J}{L_rL_x}
 \left(\Gamma-a^2p\right),
\ee
and
\be{tilde-GammaR-pz}
 \widetilde\Gamma_R
 =
 \frac{J}{L_r^2}\Gamma_R,
 \qquad
 a^2\widetilde p_z
 =
 \frac{J}{L_x^2}a^2p_z .
\ee
It follows immediately that
\be{S-projective-invariant}
 S
 \equiv
 \frac{(\Gamma-a^2p)^2}{\Gamma_Rp_z}
 =
 \frac{(\widetilde\Gamma-a^2\widetilde p)^2}
 {\widetilde\Gamma_R\widetilde p_z}
 \equiv
 \widetilde S .
\ee
Therefore the conformal structure fixed by the off-diagonal Einstein
equations is projectively invariant.  Indeed, since
\be{Sigma:SvarSigma}
 \Sigma=\epsilon\sqrt{S}\,\varSigma(y),
\ee
and since the characteristic variable \(y=R-z\) is not changed by the
M\"obius action on \(\Gamma\) and \(p\), we may take
\be{tildeSigma}
 \widetilde\Sigma=\Sigma .
\ee

The remaining structure functions must transform with the same projective
weights as \(\Gamma_R\) and \(p_z\):
\be{Delta-Y-projective-weight}
 \widetilde\Delta
 =
 \frac{J}{L_r^2}\Delta,
 \qquad
 \widetilde Y
 =
 \frac{J}{L_x^2}Y,
 \qquad
 \widetilde\Sigma=\Sigma .
\ee
Here $ Y(x)\equiv (1-x^2)Q(x).$
With these transformation laws, the orbit-space part of the metric is
preserved after the corresponding redefinition of the radial and angular
coordinates.  Since
\be{Rz-def-covariance}
 \partial_R=\Gamma'(r)\partial_r,
 \qquad
 \partial_z=\dot p(x)\partial_x,
\ee
the transformed representative is described by coordinates
\(\widetilde r\) and \(\widetilde x\) satisfying
\be{rx-projective-transform}
 \frac{d\widetilde r}{dr}
 =
 \frac{\sqrt J}{L_r},
 \qquad
 \frac{d\widetilde x}{dx}
 =
 \frac{\sqrt J}{L_x}.
\ee
Equivalently,
\be{partial-rx-projective-transform}
 \partial_{\widetilde r}
 =
 \frac{L_r}{\sqrt J}\partial_r,
 \qquad
 \partial_{\widetilde x}
 =
 \frac{L_x}{\sqrt J}\partial_x .
\ee

The Killing sector is also preserved.  Define
\be{theta-chi-def}
 \theta_p\equiv dt-ap\,d\phi,
 \qquad
 \chi_\Gamma\equiv \Gamma d\phi-a\,dt .
\ee
Together with the projective transformation \eqref{Mobius}, we perform the
linear transformation
\be{Killing-coordinate-transform}
 \begin{pmatrix}
  a\,d\widetilde t\\
  d\widetilde\phi
 \end{pmatrix}
 =
 \frac1{\sqrt J}
 \begin{pmatrix}
  A & B\\
  C & D
 \end{pmatrix}
 \begin{pmatrix}
  a\,dt\\
  d\phi
 \end{pmatrix}.
\ee
For any variable \(\lambda\) transformed as
\[
 \widetilde\lambda=\frac{A\lambda+B}{C\lambda+D},
\]
one has
\be{one-form-general-transform}
 a\,d\widetilde t-\widetilde\lambda\,d\widetilde\phi
 =
 \frac{\sqrt J}{C\lambda+D}
 \left(a\,dt-\lambda\,d\phi\right).
\ee
Choosing \(\lambda=a^2p\) and \(\lambda=\Gamma\), respectively, gives
\be{theta-transform}
 \widetilde\theta_{\widetilde p}
 \equiv
 d\widetilde t-a\widetilde p\,d\widetilde\phi
 =
 \frac{\sqrt J}{L_x}\theta_p ,
\qquad \widetilde\chi_{\widetilde\Gamma}
 \equiv
 \widetilde\Gamma d\widetilde\phi-a\,d\widetilde t
 =
 \frac{\sqrt J}{L_r}\chi_\Gamma .
\ee
Using Eqs.~\eqref{tilde-Gamma-p}, \eqref{Delta-Y-projective-weight}, and \eqref{theta-transform}, one finds
\be{Killing-sector-theta-invariant}
 -\frac{\Sigma\Delta}{(\Gamma-a^2p)^2}\theta_p^2
 =
 -\frac{\widetilde\Sigma\widetilde\Delta}
 {(\widetilde\Gamma-a^2\widetilde p)^2}
 \widetilde\theta_{\widetilde p}^{\,2},
 \quad
 \frac{\Sigma Y}{(\Gamma-a^2p)^2}\chi_\Gamma^2
 =
 \frac{\widetilde\Sigma\widetilde Y}
 {(\widetilde\Gamma-a^2\widetilde p)^2}
 \widetilde\chi_{\widetilde\Gamma}^{\,2}.
\ee
Thus the Killing part of the metric is exactly preserved.  Together with
\eqref{rx-projective-transform}, this shows that the full metric ansatz is
covariant under the common projective transformation.

The same covariance holds for the master equation.  Let
\be{master-operator}
 \mathcal E[\Delta,Y]
 \equiv
 \Delta''
 +(\partial_r\mathcal C_{\Delta'})\Delta'
 +\mathcal C_\Delta(\Delta-a^2Y)
 -(\partial_x\mathcal C_{\Delta'})\dot Y
 -\ddot Y ,
\ee
where
\be{C-functions-covariance}
 \mathcal C_{\Delta'}
 =
 \log
 \frac{\Sigma}{(\Gamma-a^2p)^3},
 \qquad
 \mathcal C_\Delta
 =
 \frac{2}{\Gamma-a^2p}
 \partial_y
 \log
 \frac{(\Gamma-a^2p)^2}{\varSigma(y)} .
\ee
Under the projective transformation, the operator written in the transformed
variables has the same form,
\bea
 \widetilde{\mathcal E}[\widetilde\Delta,\widetilde Y]
 &=&
 \widetilde\Delta_{\widetilde r\widetilde r}
 +(\partial_{\widetilde r}\widetilde{\mathcal C}_{\Delta'})
  \widetilde\Delta_{\widetilde r}
 +\widetilde{\mathcal C}_{\Delta}
  (\widetilde\Delta-a^2\widetilde Y)
 -(\partial_{\widetilde x}\widetilde{\mathcal C}_{\Delta'})
  \widetilde Y_{\widetilde x}
 -\widetilde Y_{\widetilde x\widetilde x}.
\eea
A direct substitution of
\eqref{Delta-Y-projective-weight} and
\eqref{rx-projective-transform} shows that
\be{master-equation-covariant}
 \widetilde{\mathcal E}[\widetilde\Delta,\widetilde Y]=0
 \qquad
 \Longleftrightarrow
 \qquad
 \mathcal E[\Delta,Y]=0 .
\ee
Hence the master equation is a projectively covariant equation for the
pair of structure functions \((\Delta,Y)\).

This explains why the \(\gamma=0\) representative used in the explicit
Kerr--Carter and Pleba\'nski--Demia\'nski reductions does not lose
generality.  In the exponential branch,
\be{Gamma-exp-projective}
 \Gamma(R)
 =
 \frac{\alpha e^{\mu R}+\beta e^{-\mu R}}
 {\gamma e^{\mu R}+\delta e^{-\mu R}}
 =
 \frac{\alpha w+\beta}{\gamma w+\delta},
 \qquad
 w\equiv e^{2\mu R}.
\ee
Thus \(\Gamma(R)\) itself is a M\"obius transform of \(w\).  Applying the
inverse projective transformation
\be{inverse-projective-transform}
 \widetilde\Gamma
 =
 \frac{\delta\Gamma-\beta}{-\gamma\Gamma+\alpha}
\ee
gives
\be{Gamma-to-w}
 \widetilde\Gamma=w=e^{2\mu R},
\ee
up to an irrelevant normalization and shift of \(R\).  Therefore the
denominator in the original exponential representative can be removed by a
projective gauge choice.  The \(\gamma\neq0\) sector is not a distinct local
solution branch, but a projective image of the \(\gamma=0\) representative.

\section{Compatible sources of the Carter-projective master equation} \label{sec:matter} 

In the previous sections we derived the master equation of the
Carter-projective sector of the form,
\begin{equation} \label{master2}
 \mathcal L_{\rm CP}[\Delta,Y]
 =
 16\pi\Sigma
 \left(
 T_{\hat0\hat0}+T_{\hat3\hat3}
 \right).
\end{equation}
The purpose of this section is to clarify what kind of source can be consistently coupled to the same Carter-projective geometry. 
The point is not to introduce arbitrary matter extensions of the metric ansatz, but to identify stress tensors which preserve the projective and separable structure fixed by the off-diagonal equations.

The off-diagonal equations first determine the Carter-projective sector. 
In the anti-aligned exponential branch this geometric reduction also implies the diagonal identities~\eqref{automatic-diagonal-identities}.
Therefore any source compatible with the same Carter-projective sector must satisfy
\begin{equation} \label{TT}
 T_{\hat0\hat0}+T_{\hat1\hat1}=0,
 \qquad
 T_{\hat2\hat2}-T_{\hat3\hat3}=0.
\end{equation}
Thus the automatic diagonal identities of the Kerr/Carter-projective geometry become algebraic compatibility conditions on the admissible stress tensor.

For a source diagonal in the orthonormal frame,
\begin{equation}
 T_{\hat a\hat b}
 =
 \mathrm{diag}(\rho,p_r,p_\theta,p_\phi),
\end{equation}
the compatibility conditions become
\be{compatibility}
 p_r=-\rho,
 \qquad
 p_\theta=p_\phi.
\ee
Hence a compatible diagonal source is radially vacuum-like and tangentially isotropic in the Carter-projective frame. 
After these algebraic conditions are imposed, the remaining independent matter information enters only through the combination
\begin{equation} 
T_{\hat0\hat0}+T_{\hat3\hat3} = \rho+p_\phi, 
\end{equation} 
which appears as the source term in the master equation~\eqref{master2}.

The compatibility conditions~\eqref{compatibility} are not merely technical assumptions introduced to preserve separability.
They have a direct physical interpretation.  
The first condition, $\rho+p_r=0$, is the vacuum-like radial equation of
state that appears in regular black-hole models with de Sitter-type cores
~\cite{Dymnikova1992,Hayward2006}.  
It is also natural in nonlinear electrodynamics sources for regular black holes
~\cite{AyonBeatoGarcia1998,AyonBeatoGarcia1999,Toshmatov2017} and in Kerr--Schild or Newman--Janis constructions of rotating regular geometries
~\cite{GursesGursey1975,BambiModesto2013,DymnikovaGalaktionov2015}.
The second condition, \(p_\theta=p_\phi\), states that the tangential stresses are isotropic in the Carter-projective orthonormal frame.  
Thus the Carter-projective sector does not allow an arbitrary anisotropic fluid: it selects precisely the type of stress tensor that is naturally associated with many separability-preserving rotating matter geometries.

From this viewpoint, the algebraic restrictions on the source provide a
geometric explanation for why this type of effective equation of state
frequently appears in separability-preserving regular rotating geometries. 
The condition is not imposed by hand as an independent matter ansatz; rather, it follows from requiring the matter source to be compatible with the same Carter-projective structure that underlies the Kerr--Carter and Pleba\'nski--Demia\'nski families.

The examples below should be understood in this restricted sense. 
They are not arbitrary matter deformations, but compatible sources for the Carter-projective master operator. 
In the homogeneous case, 
$
T_{\hat0\hat0}+T_{\hat3\hat3}=0, 
$
Eq.~\eqref{master2} reduces to the vacuum equation whose solutions give the Kerr--Carter and Pleba\'nski--Demia\'nski structure functions. 
For a non-vanishing compatible source, the same operator determines how the radial and angular structure functions are deformed while the underlying Carter-projective geometry is preserved.

In this sense, the off-diagonal reduction plays two roles. Geometrically, it fixes the Carter-projective sector and makes the diagonal combinations \eqref{automatic-diagonal-identities} automatic. Physically, when matter is present, the same identities restrict the allowed stress tensor. Matter fields compatible with the Carter-projective ansatz must be aligned with the separable orthonormal frame and must satisfy \eqref{TT}; their remaining freedom enters only through the source term of the master equation.

\subsection{Aligned Maxwell source}
\label{subsec:maxwell-source}

The canonical compatible source is the aligned Maxwell field.  In the
charged Pleba\'nski--Demia\'nski family the electromagnetic field is aligned
with the principal Carter-projective frame, and its stress tensor is diagonal
in the corresponding orthonormal basis.  It satisfies the compatibility
conditions
\be{Maxwell-compatibility}
 T_{\hat0\hat0}+T_{\hat1\hat1}=0,
 \qquad
 T_{\hat2\hat2}-T_{\hat3\hat3}=0,
\ee
so that the off-diagonal projective sector is unchanged.

Let 
\be{qem-def} 
q_{\rm em}^2\equiv e^2+g^2
\ee 
denote the electric--magnetic charge combination. 
In the non-accelerating Carter representative, where 
\[ \Sigma=r^2+u^2, \qquad u\equiv ax_-, \] 
the aligned Maxwell stress tensor gives 
\be{Maxwell-T-combination} 
T_{\hat0\hat0}+T_{\hat3\hat3} 
= \frac{q_{\rm em}^2}{4\pi\Sigma^2}, 
\ee 
up to the overall sign convention for the electromagnetic stress tensor and the orthonormal-frame signature. 
Therefore the source appearing in the Carter-projective master equation, \be{Maxwell-source-combination} 
\cS_{\rm em} \equiv 16\pi\Sigma \left( T_{\hat0\hat0}+T_{\hat3\hat3} \right), \ee 
is 
\be{Maxwell-source-Sigma} 
\cS_{\rm em} = \frac{4q_{\rm em}^2}{r^2+u^2}. 
\ee 
In this representative the homogeneous master operator is usually written after multiplying the equation by \(r^2+u^2\). 
Thus it is useful to define the corresponding effective source 
\be{Maxwell-effective-source} 
\mathcal J_{\rm em} \equiv (r^2+u^2)\cS_{\rm em} = 4q_{\rm em}^2 . 
\ee 
The sourced master equation then becomes 
\be{Maxwell-master-source} 
(r^2+u^2)(\Delta''-G'') -4r\Delta' +4uG' +4(\Delta-G) = \mathcal J_{\rm em}. \ee 
The Maxwell field therefore produces a constant deformation of the Carter structure functions. 
In the above convention, this constant source may be absorbed by shifting the common constant sector of the radial and angular polynomials. Equivalently, one may write 
\be{Maxwell-quartic-shift} 
\Delta(r)-G(u) = \Delta_{\rm vac}(r)-G_{\rm vac}(u) + q_{\rm em}^2 , 
\ee 
or distribute the same constant shift between \(\Delta\) and \(G\) by a redefinition of the common integration constant. 
Thus the aligned Maxwell field does not alter the projective equations or the quartic nature of the structure functions; it appears as the familiar charge deformation of the Carter--Pleba\'nski--Demia\'nski family.

This example is important because it is not an artificial matter profile.
It is the canonical compatible source of the Carter-projective sector.  
The role of the master equation is to show that the aligned Maxwell stress
tensor sources precisely the same operator that controls the vacuum
Kerr--Carter and Pleba\'nski--Demia\'nski structure functions.

\subsection{Separable anisotropic sources}
\label{subsec:monopole-source}

As a simple class of compatible sources, let us consider an anisotropic
stress tensor which is diagonal in the Carter-projective orthonormal frame,
\be{anisotropic-T}
 T_{\hat a\hat b}
 =
 \mathrm{diag}(\rho,p_r,p_\theta,p_\phi).
\ee
The compatibility conditions derived above require
$
 p_r=-\rho,$ and 
$ p_\theta=p_\phi\equiv p_\perp .$
Thus
\be{anisotropic-compatibility}
 \rho+p_r=0,
 \qquad
 p_\theta-p_\phi=0,
\ee
so that the two geometric identities
\(G_{\hat0\hat0}+G_{\hat1\hat1}=0\) and
\(G_{\hat2\hat2}-G_{\hat3\hat3}=0\) are compatible with the matter source.
The only remaining sourced diagonal combination is
\be{anisotropic-sourced-combination}
 G_{\hat0\hat0}+G_{\hat3\hat3}
 =
 8\pi(\rho+p_\perp).
\ee
Equivalently, the source entering the Carter-projective master equation is
\be{anisotropic-source-general}
 \cS(r,x)
 =
 16\pi\Sigma(\rho+p_\perp).
\ee
The Carter-projective sector selects a restricted class of anisotropic sources satisfying Eq.~\eqref{anisotropic-compatibility} rather than an arbitrary anisotropic fluid.

For concreteness, we discuss the resulting source terms in the
non-accelerating representative \(A=0\), \(\gamma=0\).  We define
\be{uG-def-anisotropic}
 u\equiv ax_-,
 \qquad
 G(u)\equiv a^2Y(x).
\ee
Then the homogeneous master equation takes the form
\be{vac-master-Kerr}
 (r^2+u^2)(\Delta''-G'')
 -4r\Delta'
 +4uG'
 +4(\Delta-G)
 =
 0 .
\ee
Compatible matter sources deform the right-hand side of this equation.
Below we describe two simple choices.

\paragraph{Constant source.}
The simplest separability-preserving choice is a constant source for the
master operator,
\be{monopole-source-profile}
 8\pi(\rho+p_\perp)
 =
 \frac{s_0}{2\Sigma},
\ee
or equivalently
\be{monopole-source-master}
 2\Sigma
 \left(
 G_{\hat0\hat0}+G_{\hat3\hat3}
 \right)
 =
 s_0 .
\ee
The master equation becomes
\be{sourced-master-simple}
 (r^2+u^2)(\Delta''-G'')
 -4r\Delta'
 +4uG'
 +4(\Delta-G)
 =
 s_0 .
\ee
It is solved by
\be{Delta-monopole}
 \Delta(r)
 =
 d_4r^4+d_2r^2+d_1r+d_0,
\ee
and
\be{Y-monopole}
 Y(x)
 =
 d_4a^2x_-^4
 -d_2x_-^2
 +\frac{g_1}{a}x_-
 +\frac{d_0}{a^2}
 -\frac{s_0}{4a^2}.
\ee
Thus the constant source preserves the Carter-projective separable
structure and appears as a constant deformation of the angular structure
function.  We refer to this as a monopole-type source in the sense that it
deforms only the constant sector of the Carter structure functions.

\paragraph{Localized radial deformation.}
A different compatible source is obtained by deforming only the radial
structure function,
\be{radial-deformation}
 \Delta(r)
 =
 \Delta_{\rm vac}(r)+f(r),
 \qquad
 G(u)=G_{\rm vac}(u),
\ee
where \(\Delta_{\rm vac}\) and \(G_{\rm vac}\) solve the homogeneous
equation \eqref{vac-master-Kerr}.  The sourced master equation then gives
\be{rho-p-source}
 8\pi(\rho+p_\perp)
 =
 \frac{(r^2+u^2)f''(r)-4r f'(r)+4f(r)}
 {2\Sigma}.
\ee
Thus the deformation function \(f(r)\) determines the effective source
combination \(\rho+p_\perp\).

To connect the deformation \(f(r)\) with the usual mass-function language,
write the vacuum radial polynomial as, writing $\Delta_1 = -2M$,
\be{Delta-vac-mass}
\Delta_{\rm vac}(r)
=\frac{r^4}{\ell^2}+\Delta_2 r^2-2Mr+\Delta_0 .
\ee
Replacing the constant mass \(M\) by an effective mass function \(m(r)\)
gives
\be{Delta-effective-mass}
\Delta(r)
=
\frac{r^4}{\ell^2}
+\Delta_2 r^2
-2m(r)r
+\Delta_0 ,
\ee
and hence
\be{f-mass-function}
f(r)
=
-2r\,[m(r)-M].
\ee
For a Hayward-type profile,
\be{Hayward-mass}
m(r)
=
\frac{Mr^3}{r^3+2ML^2},
\ee
one obtains
\be{Hayward-f}
f_{\rm H}(r)
=
\frac{4M^2L^2 r}{r^3+2M L^2}.
\ee

The corresponding source is obtained by substituting \(f_{\rm H}(r)\) into
\eqref{rho-p-source}:
\be{Hayward-source}
 8\pi(\rho+p_\perp)_{\rm H}
 =
 \frac{
 (r^2+u^2)f_{\rm H}''(r)
 -4r f_{\rm H}'(r)
 +4f_{\rm H}(r)
 }
 {2\Sigma}.
\ee
In the non-accelerating Carter representative, where
\(\Sigma=r^2+u^2\), this can be written explicitly as
\be{Hayward-source-explicit}
 8\pi(\rho+p_\perp)_{\rm H}
 =
 \frac{
 f_{\rm H}''(r)
 }{2}
 -
 \frac{
 2r f_{\rm H}'(r)-2f_{\rm H}(r)
 }{r^2+u^2}.
\ee
This example illustrates how a regular-black-hole-type radial profile can be
encoded as a compatible source for the Carter-projective master equation.
The source is not prescribed independently; it is the effective stress
combination required by the chosen radial deformation.

The examples above should not be regarded as independent fundamental matter
models.  Rather, they illustrate how compatible anisotropic stress tensors,
subject to the same algebraic conditions \eqref{anisotropic-compatibility}, enter the
Carter-projective reduction through different source profiles in the single
master equation.

\section{Conclusion} \label{sec:con}
In this work we have studied the Carter-projective sector of stationary axisymmetric spacetimes from the viewpoint of the master equation. 
The projective structure itself was taken from the off-diagonal Einstein equations. 
In this sector, \(G_{\hat0\hat3}=0\) fixes the conformal factor up to a function of the characteristic variable \(y=R-z\), while \(G_{\hat1\hat2}=0\) gives, in the generic closure branch, a common Schwarzian condition and a projective alignment between the radial and angular variables, which can be viewed as an Einstein-equation counterpart of the hidden-symmetry structure underlying the PD family.
We used this projective reduction as the input and focused on the anti-aligned exponential branch, which contains the usual Lorentzian Kerr--Carter and Pleba\'nski--Demia\'nski real section.

A key geometric feature of this branch is that, once the off-diagonal equations have fixed the Carter-projective sector, the Einstein tensor satisfies two automatic diagonal identities,
\begin{equation} 
G_{\hat0\hat0}+G_{\hat1\hat1}=0, 
\qquad 
G_{\hat2\hat2}-G_{\hat3\hat3}=0 . 
\end{equation} 
These relations are not vacuum field equations; they are geometric identities of the reduced Kerr/Carter-projective geometry. 
Consequently, when matter is included through the Einstein equation~\eqref{Einstein-eq}, they become algebraic compatibility conditions on the admissible stress tensor, $T_{\hat0\hat0}+T_{\hat1\hat1}=0$ and $ 
T_{\hat2\hat2}-T_{\hat3\hat3}=0.$ 
A noteworthy implication of the master-equation framework concerns the
matter content of regular rotating black-hole models.  
For a diagonal source in the orthonormal frame, these conditions reduces to 
\begin{equation}
p_r=-\rho,
\qquad
p_\theta=p_\phi .
\label{conc-EOS}
\end{equation}
This equation of state has long been used as a working ansatz in regular
rotating black-hole models based on de Sitter-type cores, nonlinear
electrodynamics, and Kerr--Schild constructions.  
In the present formulation it is not an independent matter postulate but a direct consequence of compatibility with the Carter-projective geometry.  
The master equation then organizes the corresponding matter sectors in a uniform way: the empirical EOS \eqref{conc-EOS} selects the admissible stress tensors, while the master equation determines how the radial and angular structure functions respond to the remaining freedom in $T_{\hat 0\hat 0}+T_{\hat 3\hat 3}$.

After these compatibility conditions are imposed, the remaining independent matter information enters through the single combination 
$
G_{\hat0\hat0}+G_{\hat3\hat3} 
= 8\pi \left( T_{\hat0\hat0}+T_{\hat3\hat3} \right). 
$
For the Carter-projective ansatz this equation becomes
\begin{equation} 
\mathcal L_{\rm CP}[\Delta,Y] 
= 16\pi\Sigma \left( T_{\hat0\hat0}+T_{\hat3\hat3} \right), 
\qquad Y(x)=(1-x^2)Q(x),
\end{equation} 
where the operator \(\mathcal L_{\rm CP}\) is determined by the projective data and the Carter-compatible conformal factor. 
This is the master equation of the Carter-projective sector.

In the homogeneous case, or whenever $T_{\hat0\hat0}+T_{\hat3\hat3}=0$,
the master equation reproduces the vacuum--\(\Lambda\) Kerr--Carter and Pleba\'nski--Demia\'nski structure functions. 
 In the non-accelerating representative it gives the Carter
quartic polynomials, while in the accelerating representative the same
operator yields the Pleba\'nski--Demia\'nski quartic structure.  
Thus the Carter and Pleba\'nski--Demia\'nski families arise as different
representatives of the same Carter-projective master equation, with the
acceleration parameter entering through the allowed conformal factor.

We have also shown that the construction is projectively covariant.
A common M\"obius transformation of the radial and angular projective variables preserves the invariant
$
 S=(\Gamma-a^2p)^2/(\Gamma_Rp_z),
$
and, with the corresponding projective weights of \(\Delta\) and \(Y\), the metric ansatz and the master equation are covariant.  
This shows that the simple representative used in the explicit calculation, such as the \(\gamma=0\) representative of the exponential branch, is a projective gauge choice rather than a loss of generality.

As discussed in Sec.~\ref{subsec:independent-diagonal}, the remaining
diagonal components are not independent after the off-diagonal projective
equations, the automatic diagonal identities, and the master equation have
been imposed.  In the homogeneous vacuum case they only fix the physical
interpretation of the integration constants, including the cosmological
constant and the normalization of the quartic coefficients.

The master equation suggests a natural way to organize matter configurations preserving Carter-projective separability. 
The aligned Maxwell field is the canonical example and corresponds to the charged Pleba\'nski--Demia\'nski family. 
More general compatible sources, such as the anisotropic examples discussed above, illustrate how the same master operator can accommodate separability-preserving matter deformations. 
In this formulation the matter sector is characterized not by an arbitrary stress tensor, but by its compatibility with the projective identities and by the separability properties of the source term \(T_{\hat0\hat0}+T_{\hat3\hat3}\).

Several directions remain open.  
In the present paper we focused on the anti-aligned exponential branch because it contains the Kerr--Carter and Pleba\'nski--Demia\'nski real section, characterized in an appropriate
representative by the sum structure
\[
 \Gamma-a^2p \sim r^2+a^2x^2 .
\]
Other projective branches should be analyzed separately.  
The aligned exponential branch corresponds to a different real section of the same complex projective structure and may lead to a difference-type combination rather than the Kerr-type sum.  
The parabolic branch \(C=0\) arises as a degeneration of the exponential representatives, while the trigonometric branch \(C=+2\mu^2\) is related by analytic continuation \(\mu\to i\mu\).
Their allowed real sections, signatures, global properties, and regularity conditions require a separate analysis.

It would also be interesting to classify the non-generic cancellation
branches of the Riccati equation, where \(F\) and \(G\) are not separately functions of \(y\) but the particular combination entering the Riccati equation becomes \(y\)-dependent for a special choice of \(\sigma(y)\).  
Such branches are expected to be overdetermined and isolated, but a systematic classification may reveal additional special solutions.  
Finally, the matter extension developed here suggests a broader
program: to classify rotating, separability-preserving matter configurations by the projective weight and separability properties of the source term in the master equation.

\section*{Acknowledgments}
This work was supported by the National Research Foundation of Korea (NRF) grant with grant numbers RS-2026-25483539 (H.K.).
AI-assisted tools were used only for language editing; all analytic derivations and scientific conclusions are the author's responsibility.

\appendix

\section{Coordinates identities} \label{app:coordinates}
We define the $R$ and $z$ coordinates as
\begin{equation}
\partial_R \equiv \Gamma'(r)\,\partial_r,
\qquad
\partial_z \equiv \dot p(x)\,\partial_x,
\qarrow 1= \dot p \partial_x z
\qarrow \dot z \dot p = 1.
\label{app:Rz:rx}
\end{equation}

Using Eq.~\eqref{app:Rz:rx}, one has the identities
\be{app:GammaR:r}
\Gamma_R\equiv \partial_R\Gamma 
= \Gamma'(r)\,\partial_r\Gamma 
= \bigl(\Gamma'\bigr)^2 , 
\quad
p_z \equiv \partial_z p = \dot p^2 \geq 0
\qarrow \dot p = \epsilon \sqrt{p_z} 
\qarrow 
\dot z =  \frac{\epsilon}{ \sqrt{p_z}},
\ee
where $\epsilon =\pm 1$.
Then, 
\be{app:Gamma'':GammaRR}
\Gamma'' = \epsilon (\sqrt{\Gamma_R})' = \frac{\epsilon}{\Gamma'} \frac{\Gamma_{RR}}{2\sqrt{\Gamma_R}} = \frac{\Gamma_{RR}}{2\Gamma_R} , \qquad
\ddot p = \epsilon \frac{d\sqrt{p_z}}{dx}
	=\frac{\epsilon}{\dot p} \frac{p_{zz}}{2 \sqrt{p_z}}
	=  \frac{p_{zz}}{2 p_z}
\ee
From these, we have, for any function  $\Sigma(r,x)$, 
\be{app:Sigma R}
\Sigma_R = \Gamma' \Sigma', \quad
\Sigma_{RR} = \Gamma'(\Gamma' \Sigma')' = \Gamma'^2 \Sigma'' + \Gamma'\Gamma'' \Sigma'
= \Gamma_R \Sigma'' +\frac{\Gamma_{RR}}{2\Gamma_R}\Sigma_R
\qarrow 
\Sigma'' = \frac{\Sigma_{RR}}{\Gamma_R} - \frac{\Gamma_{RR}}{2\Gamma_R^2} \Sigma_R .
\ee
\be{app:Sigma z}
\Sigma_z = \dot p\,\dot\Sigma, \quad
\Sigma_{zz} = p_z \ddot \Sigma+ \frac{p_{zz}}{2p_z} \Sigma_z, 
\qarrow 
\ddot \Sigma = \frac{\Sigma_{zz}}{p_z} - \frac{p_{zz}}{2p_z^2} \Sigma_z .
\ee
For example, we may choose $\Sigma=R$ or $\Sigma =z$. Then, we have the following identity:
\be{app:ddR}
R'' = - \frac{\Gamma_{RR}}{2\Gamma_R^2} , \qquad\ddot z =- \frac{p_{zz}}{2p_z^2}.
\ee

\section{Summary of the previous Carter-projective reduction} \label{app:pre}

The off-diagonal stress tensor $G_{\hat 0\hat 3}$ with  respect to the orthonormal tetrad becomes, after taking $q=\Xi=1$, 
\be{app:G03}
G_{\hat 0\hat 3} 
=-\frac{a  Q^2 \sin (\theta ) \sqrt{\Delta  Q}}
{2 \Sigma \left(\Gamma -a^2 p\right)}
   \left[-\frac{\Gamma '^2}{\Gamma -a^2p}
   	+\frac{a^2 \dot p^2}{\Gamma -a^2 p}
	+\Gamma '' 
	+\frac{\Gamma ' \Sigma '}{\Sigma }
	+\frac{\dot p \dot \Sigma}{\Sigma }
	+\ddot p\right] .
\ee
Using the characteristic coordinates, the equation $G_{\hat 0\hat 3} = 0$ determines the functional form of $\Sigma(R,z)$, 
\be{app:Sigma}
\Sigma(R,z) = \frac{(\Gamma -a^2p) \varSigma(y)}{\sqrt{\Gamma_R p_z}}, \qquad y \equiv R-z, \qquad 
\bar y \equiv R + z.
\ee

The $G_{\hat 1\hat 2}=0$ becomes
\be{app:G12}
\left( \frac{a^2 
   \dot {p}}{\Gamma -a^2 p}\right)'
   =\frac{\dot{\Sigma}\Sigma' }{\Sigma^2}
   	-\frac23 \frac{\dot{\Sigma}' }{\Sigma} .
\ee
Writing the equation in terms of $R$ and $z$ coordinates and using $y\equiv R-z$, the equation is reduced to the Riccati type,
\be{app:Riccati}
\mathring \sigma = \frac12 \sigma^2 + \frac12 F(R,z) \sigma + \frac14G(R,z) , 
\ee
where 
\bea
F(R,z) &=&   
	\frac{\Gamma_R+a^2p_z}{\Gamma-a^2p}
		+\frac12\,\partial_z\log p_z	
		-\frac12\,\partial_R\log\Gamma_R
	 , \nn \\
G(R,z)&=& -\left[
\frac{\Gamma_{RR}}{\Gamma_R} \frac{a^2p_z}{\Gamma-a^2p}
	-\frac{p_{zz}}{p_z}\frac{\Gamma_R}{\Gamma-a^2p}
	+\frac1{2}\,\frac{\Gamma_{RR}}{\Gamma_R} 
		\frac{p_{zz}}{p_z}
\right] .
\eea

In the generic branch, the Riccati equation is an ordinary differential equation in \(y\). 
Hence we require
$$
F=F(y),
\qquad
G=G(y).
$$
Then, from $F= \partial_y \log S$, the condition $F\equiv F(y)$ implies 
 $$
 \partial_{\bar y} \partial_y \log S =0
 $$ 
which leads to the equality of the Schwarzian derivatives, 
\be{app:SchD}
\{\Gamma, R\}= \{p, z\} = C, \qquad \{A,B\}\equiv \frac{A_{BBB}}{A_B}-\frac{3}{2}\left(\frac{A_{BB}}{A_B}\right)^2.
\ee

\vspace{.3cm}
\paragraph{Remark on a non-generic cancellation branch.}
In the main text, we have focused on the generic closure condition in
which both $F$ and $G$ are functions of the characteristic variable
$y=R-z$ separately. There is, however, a logically possible non-generic
alternative. Even when $F(R,z)$ and $G(R,z)$ are not separately functions of
$y$, the particular combination appearing in the Riccati equation may become
a function of $y$ for a special choice of $\sigma(y)$. More explicitly, the
Riccati equation can be written as
\begin{equation}
 \mathring\sigma-\frac12\sigma^2
 =
 \frac12 F(R,z)\sigma(y)+\frac14 G(R,z) .
\end{equation}
Since the left-hand side is a function of $y$ only, the right-hand side must
also be independent of $\bar y=R+z$. Therefore, if
$\partial_{\bar y}F\neq0$, a necessary condition is
\begin{equation}
 \sigma_*(R,z)
 =
 -\frac12
 \frac{\partial_{\bar y}G(R,z)}
      {\partial_{\bar y}F(R,z)} .
 \label{sigma:special}
\end{equation}
For this to define an admissible branch, the quantity obtained in
\eqref{sigma:special} must itself be a function of $y$ only,
\begin{equation}
 \partial_{\bar y}\sigma_*=0,
 \label{special-cond-1}
\end{equation}
and it must satisfy the original Riccati equation,
\begin{equation}
 \mathring\sigma_*
 -\frac12\sigma_*^2
 -\frac12F(R,z)\sigma_*
 -\frac14G(R,z)
 =0 .
 \label{special-cond-2}
\end{equation}
These conditions are overdetermined for generic choices of
$\Gamma(R)$ and $p(z)$. We therefore regard this possibility as a
non-generic cancellation branch. In the present work we restrict attention
to the generic closure branch, for which
\begin{equation}
 F=F(y),
 \qquad
 G=G(y),
\end{equation}
and leave the possible isolated solutions of
\eqref{special-cond-1}--\eqref{special-cond-2} for separate investigation.

\section{Checks for the anti-aligned exponential branch}
\label{app:checks}

In this appendix we verify that the anti-aligned exponential branch used in
the main text satisfies the two off-diagonal Einstein equations.  We work in
the gauge \(q=1=\Xi\).

\subsection{The equation \(G_{\hat0\hat3}=0\)}
\label{app:G03-check}

For the metric \eqref{metric:gen2}, the equation \(G_{\hat0\hat3}=0\)
takes the form
\be{G03-check-eq}
(\partial_R+\partial_z)
\log
\frac{\Sigma\sqrt{\Gamma_Rp_z}}{\Gamma-a^2p}
=0 .
\ee
Thus the quantity inside the logarithm must be a function of
\(y=R-z\) only.

For the anti-aligned exponential branch, using the forms for $\Gamma(R)$, $p(z)$ in Eq.~\eqref{GammaR:anti}, we get
\be{app-Gamma-minus-p}
\Gamma-a^2p
=
-\frac{D_0
\left(
e^{\mu(R+z_\ominus)}
+
e^{-\mu(R+z_\ominus)}
\right)}
{
\left(\gamma e^{\mu R}+\delta e^{-\mu R}\right)
\left(\gamma e^{-\mu z_\ominus}
-\delta e^{\mu z_\ominus}\right)
}.
\ee
By using this result and from Eq.~\eqref{GammaR-pz-anti}, direct calculation gives 
\be{app-S-cosh}
S
\equiv
\frac{(\Gamma-a^2p)^2}{\Gamma_Rp_z}
=
\frac{a^2}{\mu^2}
\cosh^2[\mu(\bar y-z_r)] .
\ee
With the solution~\eqref{varSigma} for the Riccati equation, the conformal factor becomes
\be{app-Sigma-anti}
\Sigma
=
\epsilon\sqrt{S}\,\varSigma(y)
=
\frac{\epsilon a\varSigma_0}{\mu}
\frac{\cosh[\mu(\bar y-z_r)]}
{\left(
Ae^{\mu(y+z_r)/2}
+
Be^{-\mu(y+z_r)/2}
\right)^2}.
\ee
Hence
\be{app-G03-argument}
\frac{\Sigma\sqrt{\Gamma_Rp_z}}{\Gamma-a^2p}
=
\frac{\Sigma}{\epsilon\sqrt{S}}
=
\varSigma(y).
\ee
Since this is a function of \(y\) only, Eq.~\eqref{G03-check-eq} is
satisfied.

\subsection{The equation \(G_{\hat1\hat2}=0\)}
\label{app:G12-check}

In the same gauge, the equation \(G_{\hat1\hat2}=0\) may be written as
\be{app:G12}
\left(
\frac{a^2p_z}{\Gamma-a^2p}
\right)_R
=
\frac13
\left[
\frac{\Sigma_R\Sigma_z}{\Sigma^2}
-
2\partial_R\partial_z\log\Sigma
\right].
\ee
Using the definition of \(S\), the left-hand side is
\be{app-G12-LHS}
\left(
\frac{a^2p_z}{\Gamma-a^2p}
\right)_R
=
-\frac{a^2p_z\Gamma_R}{(\Gamma-a^2p)^2}
=
-\frac{a^2}{S}
=
-\frac{\mu^2}{\cosh^2[\mu(\bar y-z_r)]}.
\ee

On the other hand, since
\[
\Sigma=\epsilon\sqrt{S(\bar y)}\,\varSigma(y),
\]
we have
\bea
\partial_R\log\Sigma
&=&
\frac12\frac{S_{\bar y}}{S}
+
\frac{\varSigma_y}{\varSigma},
\nn\\
\partial_z\log\Sigma
&=&
\frac12\frac{S_{\bar y}}{S}
-
\frac{\varSigma_y}{\varSigma},
\nn\\
\partial_R\partial_z\log\Sigma
&=&
\frac12
\left(
\frac{S_{\bar y}}{S}
\right)_{\bar y}
-
\left(
\frac{\varSigma_y}{\varSigma}
\right)_y .
\label{app-logSigma-identities}
\eea
Therefore the right-hand side of Eq.~\eqref{app:G12} becomes
\be{app-G12-RHS-general}
\frac13
\left[
\frac14
\left(
\frac{S_{\bar y}}{S}
\right)^2
-
\left(
\frac{S_{\bar y}}{S}
\right)_{\bar y}
-
\left(
\frac{\varSigma_y}{\varSigma}
\right)^2
+
2
\left(
\frac{\varSigma_y}{\varSigma}
\right)_y
\right].
\ee

For the branch under consideration,
\be{app-useful-identities}
2
\left(
\frac{\varSigma_y}{\varSigma}
\right)_y
-
\left(
\frac{\varSigma_y}{\varSigma}
\right)^2
=
-\mu^2,
\qquad
\frac{S_{\bar y}}{S}
=
2\mu\tanh[\mu(\bar y-z_r)] .
\ee
It follows that
\bea
\frac13
\left[
\frac14
\left(
\frac{S_{\bar y}}{S}
\right)^2
-
\left(
\frac{S_{\bar y}}{S}
\right)_{\bar y}
-\mu^2
\right]
&=&
-\frac{\mu^2}{\cosh^2[\mu(\bar y-z_r)]}.
\eea
This agrees with Eq.~\eqref{app-G12-LHS}.  Hence
\(G_{\hat1\hat2}=0\) is also satisfied.

\section{Coordinate representatives}
\label{app:coordinates2}

In this appendix we summarize the relation between the projective
coordinates \((R,z)\) and the coordinates \((r,x)\) used in the metric
representative.  The relations follow by integrating Eq.~\eqref{Rz:rx}.
For the anti-aligned exponential branch, $\Gamma_R$ and $p_z$ is given in Eq.~\eqref{GammaR-pz-anti}.
We assume \(\mu>0\), \(\gamma\geq0\), and choose the real patch in which \(\mu D_0/a^2>0\).

The radial coordinate \(r\) is obtained from
\[
 \frac{dR}
 {\left|\gamma e^{\mu R}+\delta e^{-\mu R}\right|}
 =
 \frac{dr}{\epsilon\sqrt{2\mu D_0}},
\]
where \(\epsilon=\pm1\).  Defining
\[
 \kappa\equiv
 \sqrt{\left|\frac{\gamma\delta\mu}{2D_0}\right|},
 \qquad
 R_0\equiv
 \frac{1}{2\mu}\log\frac{|\delta|}{\gamma},
\]
one obtains the following local representatives:
\be{R:r-summary}
e^{\mu R}
=
\epsilon
\begin{cases}
\sqrt{\dfrac{\delta}{\gamma}}\,
\tan\!\left[\kappa(r-r_0)\right],
& \delta>0,
\\[8pt]
\sqrt{\dfrac{-2\beta}{\mu\gamma}}\,
\dfrac1{r-r_0},
& \delta=0,
\\[10pt]
\sqrt{\dfrac{|\delta|}{\gamma}}\,
\tanh\!\left[\kappa(r-r_0)\right],
& \delta<0,\quad R<R_0,
\\[8pt]
\sqrt{\dfrac{|\delta|}{\gamma}}\,
\coth\!\left[\kappa(r-r_0)\right],
& \delta<0,\quad R>R_0,
\\[8pt]
-\sqrt{\dfrac{\mu\delta}{2\alpha}}\,(r-r_0),
& \gamma=0.
\end{cases}
\ee
Correspondingly,
\be{Gamma-r-summary}
\Gamma(r)
=
\Gamma_0+
\begin{cases}
-\dfrac{D_0}{\delta\gamma}
\cos^2\!\left[\kappa(r-r_0)\right],
& \delta>0,
\\[8pt]
-\dfrac{\mu}{2}(r-r_0)^2,
& \delta=0,
\\[8pt]
-\dfrac{D_0}{\gamma\delta}
\cosh^2\!\left[\kappa(r-r_0)\right],
& \delta<0,\quad R<R_0,
\\[8pt]
\dfrac{D_0}{\gamma\delta}
\sinh^2\!\left[\kappa(r-r_0)\right],
& \delta<0,\quad R>R_0,
\\[8pt]
\dfrac{\mu}{2}(r-r_0)^2,
& \gamma=0,
\end{cases}
\ee
where
\[
 \Gamma_0=
 \begin{cases}
 \dfrac{\alpha}{\gamma}, & \gamma\neq0,\\[6pt]
 \dfrac{\beta}{\delta}, & \gamma=0 .
 \end{cases}
\]

Similarly, the angular coordinate \(x\) is obtained from
\[
 \frac{dz}
 {\gamma e^{-\mu z_\ominus}-\delta e^{\mu z_\ominus}}
 =
 \frac{dx}{\epsilon\sqrt{2\mu D_0/a^2}} .
\]
The corresponding representatives are
\be{z:x-summary}
e^{\mu z_\ominus}
=
\epsilon
\begin{cases}
\sqrt{\dfrac{\gamma}{\delta}}\,
\tanh\!\left[a\kappa(x-x_0)\right],
& \delta>0,\quad
e^{\mu z_\ominus}<\sqrt{\gamma/\delta},
\\[8pt]
\sqrt{\dfrac{\gamma}{\delta}}\,
\coth\!\left[a\kappa(x-x_0)\right],
& \delta>0,\quad
e^{\mu z_\ominus}>\sqrt{\gamma/\delta},
\\[8pt]
a\sqrt{\dfrac{\gamma\mu}{-2\beta}}\,(x-x_0),
& \delta=0,
\\[8pt]
\sqrt{\dfrac{\gamma}{|\delta|}}\,
\tan\!\left[a\kappa(x-x_0)\right],
& \delta<0,
\\[8pt]
\sqrt{\dfrac{2\alpha}{\delta\mu}}\,
\dfrac1{a(x-x_0)},
& \gamma=0.
\end{cases}
\ee
and
\be{p-x-summary}
a^2p(x)
=
\Gamma_0+
\begin{cases}
\dfrac{D_0}{\gamma\delta}
\sinh^2\!\left[a\kappa(x-x_0)\right],
& \delta>0,\quad
e^{\mu z_\ominus}<\sqrt{\gamma/\delta},
\\[8pt]
-\dfrac{D_0}{\gamma\delta}
\cosh^2\!\left[a\kappa(x-x_0)\right],
& \delta>0,\quad
e^{\mu z_\ominus}>\sqrt{\gamma/\delta},
\\[8pt]
\dfrac{\mu}{2}a^2(x-x_0)^2,
& \delta=0,
\\[8pt]
-\dfrac{D_0}{\gamma\delta}
\sin^2\!\left[a\kappa(x-x_0)\right],
& \delta<0,
\\[8pt]
-\dfrac{\mu}{2}a^2(x-x_0)^2,
& \gamma=0.
\end{cases}
\ee

The representative used in the main text is the \(\gamma=0\) one.  After
choosing \(r_0=0\), shifting \(x\) by \(x_0\), and absorbing irrelevant
signs into \(r\) and \(x_-\equiv x-x_0\), it reduces to
\be{main-representative-coordinates}
e^{\mu R}
=
\epsilon
\sqrt{\frac{\mu\delta}{2\alpha}}\,r,
\qquad
e^{\mu z_\ominus}
=
\epsilon
\sqrt{\frac{2\alpha}{\delta\mu}}\,
\frac1{a x_-},
\ee
and
\be{main-representative-Gamma-p}
\Gamma(r)
=
\Gamma_0+\frac{\mu}{2}r^2,
\qquad
a^2p(x)
=
\Gamma_0-\frac{\mu}{2}a^2x_-^2 .
\ee
Thus
\be{main-representative-sum}
\Gamma-a^2p
=
\frac{\mu}{2}
\left(r^2+a^2x_-^2\right),
\ee
which is the Kerr--Carter sum structure used throughout the main text.

\section{Automatic diagonal equations}
\label{app:diagonal-eqs}

In this appendix we show that, once the anti-aligned exponential branch is
imposed, the two diagonal combinations
\[
 G_{\hat0\hat0}+G_{\hat1\hat1},
 \qquad
 G_{\hat2\hat2}-G_{\hat3\hat3}
\]
vanish identically.  We work in the representative \(q=1=\Xi\).

\subsection{The combination \(G_{\hat0\hat0}+G_{\hat1\hat1}\)}
\label{app:G00G11}

For the metric \eqref{metric:gen2}, one finds
\be{G00+G11-app}
G_{\hat0\hat0}+G_{\hat1\hat1}
=
\frac{\Delta}{2\Sigma}\,\mathcal A ,
\ee
where
\be{A-def-app}
\mathcal A
\equiv
\frac{a^2\dot p^{\,2}}{(\Gamma-a^2p)^2}
+3\frac{\Sigma'^2}{\Sigma^2}
-2\frac{\Sigma''}{\Sigma}
-\frac{2\Gamma'}{\Gamma-a^2p}\frac{\Sigma'}{\Sigma}.
\ee
Using the formula~\eqref{app:Sigma R}, this expression can be rewritten as
\be{A-R-form}
\Gamma_R\mathcal A
=
\frac{a^2}{S}
-2\left(\frac{\Sigma_R}{\Sigma}\right)_R
+\frac{\Sigma_R}{\Sigma}
\left[
 \frac{\Gamma_{RR}}{\Gamma_R}
 -\frac{2\Gamma_R}{\Gamma-a^2p}
\right].
\ee
Here we have used Eq.~\eqref{S:Sigma}.

For the anti-aligned exponential branch,
\[
 \Sigma=\epsilon\sqrt{S(\bar y)}\,\varSigma(y),
 \qquad
 S=\frac{a^2}{\mu^2}\cosh^2[\mu(\bar y-z_r)] .
\]
Moreover,
\be{Gamma-identity-app}
\frac{\Gamma_{RR}}{\Gamma_R}
-\frac{2\Gamma_R}{\Gamma-a^2p}
=
-\,\frac{S_{\bar y}}{S}.
\ee
Therefore Eq.~\eqref{A-R-form} becomes
\be{A-identity-app}
\Gamma_R\mathcal A
=
\frac{a^2}{S}
-\left(\frac{S_{\bar y}}{S}\right)_{\bar y}
-\frac14
\left(\frac{S_{\bar y}}{S}\right)^2
+\left(\frac{\varSigma_y}{\varSigma}\right)^2
-2\left(\frac{\varSigma_y}{\varSigma}\right)_y .
\ee
The two elementary identities
\be{S-varSigma-identities-app}
\frac{S_{\bar y}}{S}
=
2\mu\tanh[\mu(\bar y-z_r)],
\qquad
2\left(\frac{\varSigma_y}{\varSigma}\right)_y
-
\left(\frac{\varSigma_y}{\varSigma}\right)^2
=
-\mu^2
\ee
then give
\[
\Gamma_R\mathcal A
=
\frac{\mu^2}{\cosh^2[\mu(\bar y-z_r)]}
-\mu^2
\left[
1-\tanh^2[\mu(\bar y-z_r)]
\right]
=0 .
\]
Hence
\[
 G_{\hat0\hat0}+G_{\hat1\hat1}=0 .
\]

\subsection{The combination \(G_{\hat2\hat2}-G_{\hat3\hat3}\)}
\label{app:G22G33}

The second diagonal combination is
\be{G22-G33-app}
G_{\hat2\hat2}-G_{\hat3\hat3}
=
\frac{Q\sin^2\theta}{2\Sigma}\,\mathcal B ,
\ee
with
\be{B-def-app}
\mathcal B
\equiv
-2\frac{\ddot\Sigma}{\Sigma}
+3\frac{\dot\Sigma^2}{\Sigma^2}
+\frac{2a^2\dot p}{\Gamma-a^2p}
 \frac{\dot\Sigma}{\Sigma}
+\frac{a^2\Gamma'^2}{(\Gamma-a^2p)^2}.
\ee
In terms of the projective coordinate \(z\), this expression can be written
as
\be{B-z-form}
p_z\mathcal B
=
\frac{a^2}{S}
-\left(\frac{S_{\bar y}}{S}\right)_{\bar y}
-\frac14
\left(\frac{S_{\bar y}}{S}\right)^2
+\left(\frac{\varSigma_y}{\varSigma}\right)^2
-2\left(\frac{\varSigma_y}{\varSigma}\right)_y .
\ee
The right-hand side is the same combination that appeared in
\eqref{A-identity-app}.  Therefore, using the identities
\eqref{S-varSigma-identities-app}, we immediately obtain
\[
 p_z\mathcal B=0 .
\]
Thus
\[
 G_{\hat2\hat2}-G_{\hat3\hat3}=0 .
\]

Consequently, for the anti-aligned exponential branch, the two diagonal
combinations above are automatically satisfied.  After the off-diagonal
equations have fixed the projective sector, the only remaining independent
diagonal equation is the master equation derived from
\(G_{\hat0\hat0}+G_{\hat3\hat3}=0\).

\section{Derivation of the master equation}
\label{app:master}

In this appendix we derive the master equation used in the main text.
We work in the gauge \(q=1=\Xi\) and use the metric
\eqref{metric:gen2}.  The independent remaining diagonal combination is
\(G_{\hat0\hat0}+G_{\hat3\hat3}\).

A direct computation gives
\be{App:G00+G33}
2\Sigma
\left(
G_{\hat0\hat0}+G_{\hat3\hat3}
\right)
=
\Delta''
+
\left(
\log\frac{\Sigma}{(\Gamma-a^2p)^3}
\right)'
\Delta'
+
\mathcal C_\Delta \Delta
+
\mathcal R_{\Delta^0}
+
X(x),
\ee
where
\be{CDelta-app}
\mathcal C_\Delta
=
\frac{2\Gamma'}{\Gamma-a^2p}
\left[
\frac{2\Gamma'}{\Gamma-a^2p}
-\frac{\Sigma'}{\Sigma}
-\frac{\Gamma''}{\Gamma'}
\right]
+
\frac{2a^2\dot p^{\,2}}{(\Gamma-a^2p)^2}.
\ee
Here a prime denotes \(d/dr\), while a dot denotes \(d/dx\).
The term \(X(x)\) is
\be{X-app}
X(x)
=
Q
\left[
2+4x\frac{\dot Q}{Q}
-(1-x^2)\frac{\ddot Q}{Q}
\right].
\ee
Introducing the angular structure function, $Y(x)$, one immediately finds
\be{X-Y-app}
X(x)=-\ddot Y(x).
\ee

We next rewrite the coefficient \(\mathcal C_\Delta\) using the
Carter-projective form of the conformal factor $\Sigma$ in Eq.~\eqref{Sigma-S}.
Using
$
\partial_y=\dfrac12(\partial_R-\partial_z),
$
one obtains
\be{CDelta-final-app}
\mathcal C_\Delta
=
\frac{2}{\Gamma-a^2p}
\partial_y
\log
\frac{(\Gamma-a^2p)^2}{\varSigma(y)} .
\ee
We also define
\be{CDeltaprime-app}
\mathcal C_{\Delta'}
\equiv
\log
\frac{\Sigma}{(\Gamma-a^2p)^3}.
\ee
Then the first two terms in \eqref{App:G00+G33} are simply
\[
\Delta''
+
(\partial_r\mathcal C_{\Delta'})\Delta' .
\]

It remains to organize the terms in \(\mathcal R_{\Delta^0}\).  After
writing all angular derivatives in terms of \(Y=(1-x^2)Q\), the terms
linear in \(Y\) and \(\dot Y\) take the form
\be{RDelta-summary-app}
\mathcal R_{\Delta^0}
=
\widetilde B(r,x)Y(x)
-
(\partial_x\mathcal C_{\Delta'})\dot Y(x),
\ee
where
\be{Btilde-app}
\widetilde B
=
\left(
\log
\frac{\sqrt{p_z}\Sigma}{(\Gamma-a^2p)^2}
\right)_x
\left[
\log(\Gamma-a^2p)^2
\right]_x
-
\frac{2a^2\Gamma_R}{(\Gamma-a^2p)^2}.
\ee
Using again the conformal factor~\eqref{Sigma-S}, this coefficient reduces to
\be{Btilde-CDelta-app}
\widetilde B = -a^2\mathcal C_\Delta .
\ee
Therefore
\be{RDelta-final-app}
\mathcal R_{\Delta^0}
= -a^2\mathcal C_\Delta Y-(\partial_x\mathcal C_{\Delta'})\dot Y .
\ee

Substituting \eqref{X-Y-app} and \eqref{RDelta-final-app} into
\eqref{App:G00+G33}, we obtain
\be{master-app-final}
2\Sigma
\left(
G_{\hat0\hat0}+G_{\hat3\hat3}
\right)
=
\Delta''
+
(\partial_r\mathcal C_{\Delta'})\Delta'
+
\mathcal C_\Delta(\Delta-a^2Y)
-
(\partial_x\mathcal C_{\Delta'})\dot Y
-
\ddot Y .
\ee
Thus the Einstein equation
$
G_{\hat0\hat0}+G_{\hat3\hat3}= 8 \pi (T_{\hat 0\hat 0} + T_{\hat 3\hat 3}) 
$
is equivalent to the master equation
\be{master-equation-app}
\Delta''
+
(\partial_r\mathcal C_{\Delta'})\Delta'
+
\mathcal C_\Delta(\Delta-a^2Y)
-
(\partial_x\mathcal C_{\Delta'})\dot Y
-
\ddot Y
= \cS,
\ee
where 
$$
\cS(r,x) \equiv 16\pi\Sigma (T_{\hat 0\hat 0} + T_{\hat 3\hat 3}) .
$$


\end{document}